# Unexpected fading of comet C/2003 T4 (LINEAR) and disintegration of C/2012 S1 (ISON)

Jakub Cerny

*Society for interplanetary matter, Kraví hora 522/2, Brno-střed, 616 00, Czech Republic*
*Version August 17, 2014*

## Abstract

*Comet C/2003 T4 (LINEAR) exhibit a large asymmetry in brightness before and after perihelion, when it became much fainter. Large non-gravitational forces shows that mass of nucleus doesn't exceed $2.51 \times 10^{11}$ kg, which is nearly double than previously disintegrated comets C/2012 S1 (ISON) and C/1999 S4 (LINEAR). Amount of water mass loss of C/2003 T4 in interval between 70 days before to 60 after perihelion, was nearly $3.16(+/-0.60) \times 10^{10}$ kg or >13 % of total nucleus mass and >21 % for C/2012 S1, which is much larger than value >7 % previously stated for C/1999 S4.*

## 1. Introduction

Many newly discovered comets that was observed in past survived their perihelion passage while some others disintegrated totally, however there is comet that potentially lies on a border line between surviving and disintegrating comets. Comet C/2003 T4 (LINEAR) was discovered 13. October 2003 by Lincoln Laboratory Near-Earth Asteroid Research project. Its original semimajor axis before entering inner part of solar system was ~2521.6 AU, and the comet was moving on orbit with period >120 k years, thus it was not a dynamically new[1]. After its discovery, its brightness starts to lack behind expectations and comet continue with very decent brightening trend until perihelion. Nearly 60 days after perihelion, it become practically unobservable. Later, despite improving observing condition in southern hemisphere, it wasn't observed anymore, except astrometric observations, which magnitude data suggests that meanwhile, this comet unexpectedly fade. Precise orbit calculated for this comet exhibit large non-gravitational forces pointing to small and very active nucleus, that could be heavily damaged during its perihelion passage.

Comet C/2012 S1 (ISON) was discovered 21. September 2012 by Vitali Nevski and Artyom Novichonok, using International Scientific Optical Network (ISON) telescope. It is first known dynamically new comet on sungrazing orbit. This comet also lack in its brightness evolution while it was approaching to Sun, after extremely chaotic evolution of its brightness, its entire nucleus has disintegrated 3.5 hours before reaching its perihelion[2]. It exhibit large non-gravitational forces in its movement.

Known non-gravitational forces and water production rates allows us to calculate maximal limit for cometary nuclei mass. For previously disintegrated comet C/1999 S4 (LINEAR), Z. Sekanina calculated its maximal nucleus mass to be $1.0 \times 10^{11}$ Kg with >7 % fraction of water, calculated from its production rate before disintegration. M - the mass of cometary nucleus, can be derived from the law of momentum conservation by equation:

$$M = k_0 \, v_{H2O} \, M_{H2O} / J_r$$

Where $k_0$ is the coefficient describing degree of spatial distribution for water molecular flow which can be positive number $\leq 1$, where 1 is for 100% sunward gas flow. $v_{H2O}$ is speed of sublimating water molecules and $M_{H2O}$ is mass of water molecules produced in 1 second. $J_r$ is the radial component of non-gravitational acceleration. As the $k_0$ coefficient is unknown, but always $\leq 1$, the real nucleus mass must be always equal or smaller than resulting mass using $k_0 = 1$ in equation.[3]

The water production rate can be measured using various patterns and it also can be estimated using correlation between visual magnitudes and water production rates introduced by L. Jorda, J. Crovisier and D.W.E. Green:

log Q [H2O] = 30.675 - 0.2453 $m_H$

where $m_H$ is the heliocentric visual magnitude[4].

## 2. Analysis

### 2.1. Light curve analysis

Brightness evolution of comet C/2003 T4 can be determined on interval 550 days before and 60 days after perihelion, using photometric magnitudes from publicly available sources[5][6]. The last astrometric observation was done almost 324 days after perihelion, unfortunately in interval after 60 days after perihelion, there are missing photometric data and only astrometric magnitudes from MPC are available[1]. As its seen on Figure 1., in well covered interval, the MPC magnitudes shows extreme low quality and large majority of them are erroneous. However in distances larger than 4 AU the scatter is smaller, as the comet distance was larger, therefore coma was smaller and more condensed. Magnitude of comet can be described by standard equation:

$m_1$ = $H_{11}$ + 5 log(D) + 2.5 n log(r)

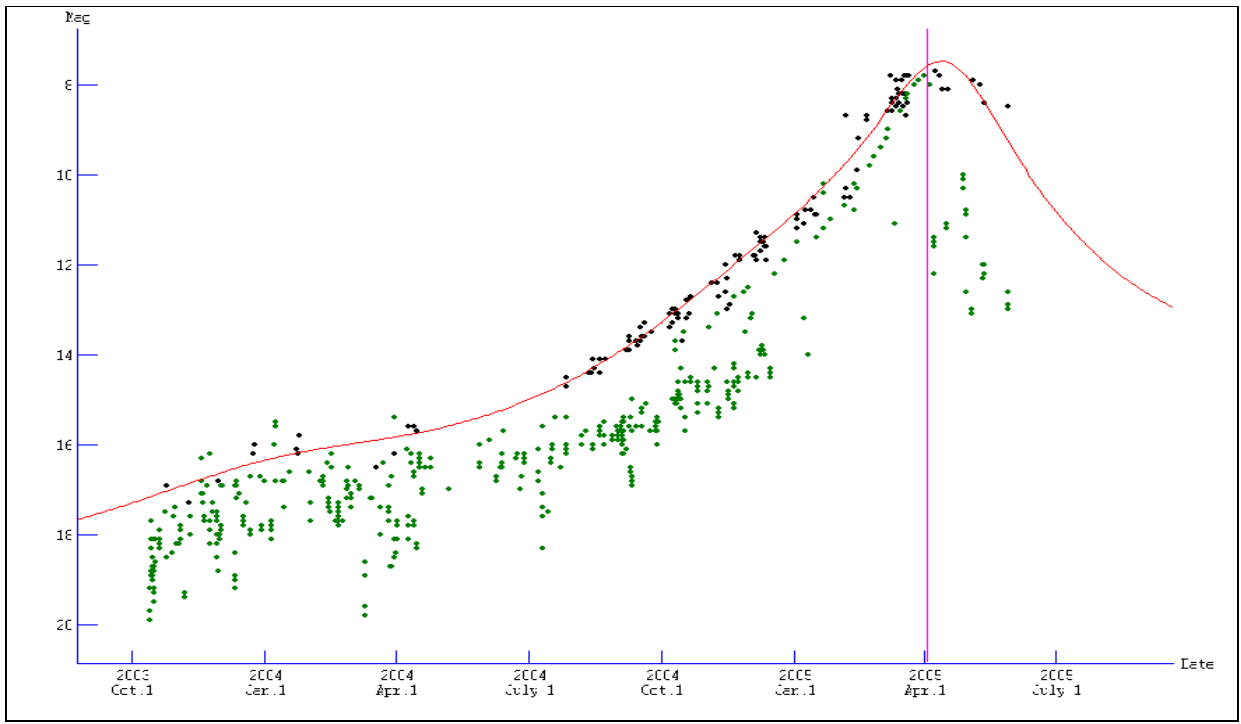

**Figure 1.** Plot shows photometric data (black dots) against astrometric MPC magnitudes (green dots). It is noticeable, that the extreme scatter of MPC data prevents them from any further detailed analysis, because of their poor quality.

Where $H_{11}$ is comet (absolute) magnitude at distance 1AU from both Earth and Sun and n is brightening factor. Photometric magnitude well fits values $H_{11}$ = 7.75 mag and n = 2.66 in whole interval 550 days before to 60 days after perihelion. The factor n corresponds to extremely slow brightening as the value 2 is adopted for inactive body without cometary activity, and 4 for ordinary comets. The further evolution is unknown, but the astrometric magnitudes indicating much smaller brightness then it would be expected. Using fixed value n = 2.66 for MPC astrometric magnitudes, outside 4 AU distance, indicating 2.15 mag drop in absolute magnitude after perihelion passage, therefore large drop in total nucleus activity. On other hand, analysing light curve of comet C/2012 S1 (ISON) on figure 3. shows much more complicated behaviour, which can't be described by one simple formula.

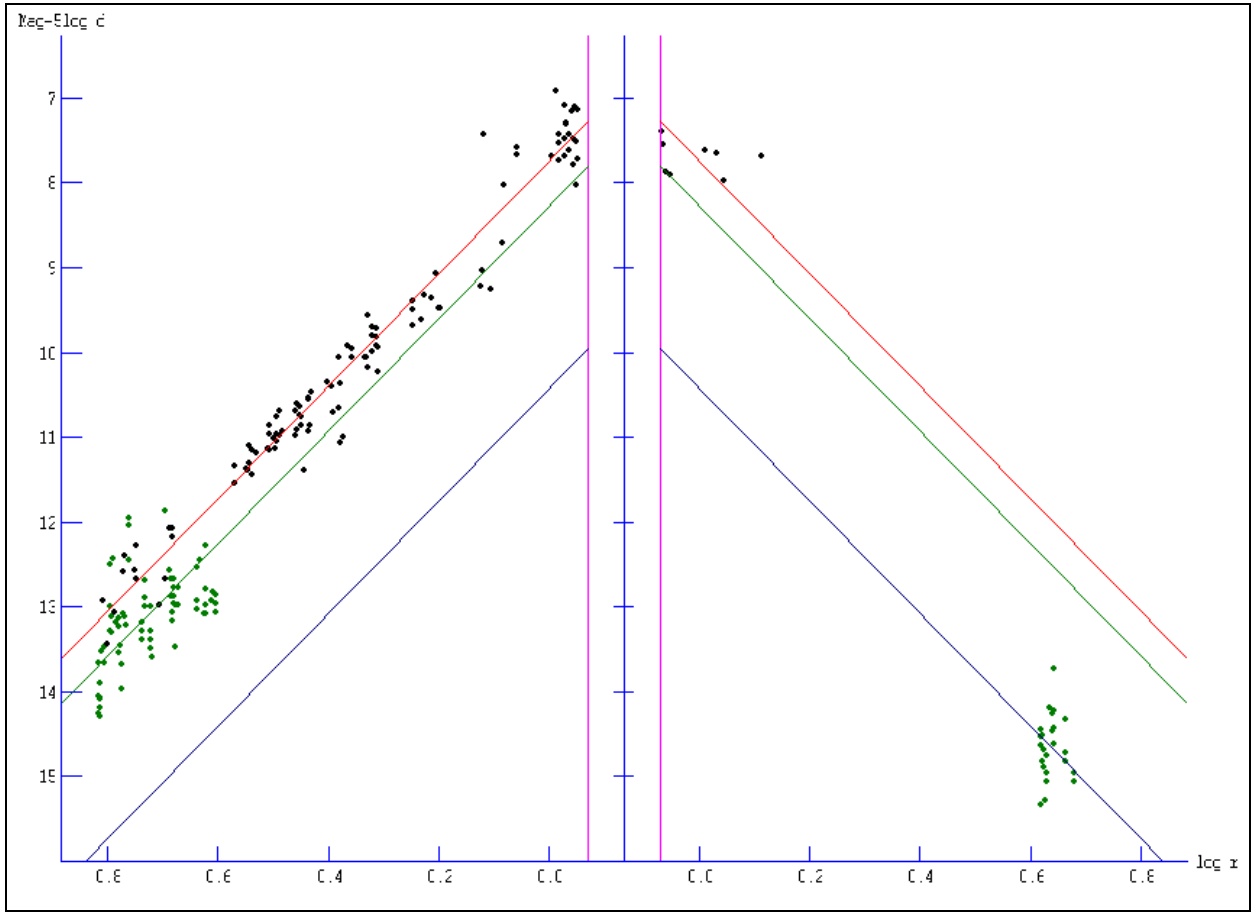

**Figure 2.** Plot of pre and post perihelion brightness evolution. Red line is the calculated brightness evolution from photometric data, green line corresponds to MPC astrometric magnitudes in distances larger than 4 AU pre-perihelion using constant n = 2.66. Blue line is for astrometric magnitudes post-perihelion in distances larger than 4 AU using constant n = 2.66. Difference pre and post perihelion MPC magnitude is as large as 2.15 mag .

### 2.2. C/2003 T4 (LINEAR) nucleus mass and water loss

Based on the absolute magnitude $H_{11}$ and empiric formula in section 1. the resulting water production rate in 1 AU distance from Sun is $5.94 \times 10^{28}$ $H_2O$ molecules per second, which is equal to $M_{H2O}$ $1.78 \times 10^3$ kg/s. The radial component of non-gravitational acceleration at 1 AU is designated $A_1$ and stated by JPL as $2.24 \times 10-4$ cm/s$^2$[7] and $v_{H2O}$ at 1 AU was estimated as 3 cm/s (value used for C/1999 S4 by Sekanina)[1]. Resulting limit for maximal mass of C/2003 T4 nucleus is $2.38 \times 10^{11}$ kg.

The amount of water produced near heliocentric distance 1 AU (1.5 AU to q) until end of photometric observations (T+60) based on calculated photometric parameters in section 2. $3.16(+/- 0.60) \times 10^{10}$ kg or >13 % of nuclear mass fraction.

### 2.3. C/2012 S1 (ISON) nucleus mass and water loss

For comet ISON, water production rate is well covered by SWAN instrument on SOHO spacecraft. The measured water production rate at 1 AU is $2.50 \times 10^{28}$ H$_2$O molecules per second[8], which is equal to $M_{H2O}$ $0.75 \times 10^3$ kg/s. $A_1$ calculated by JPL as $1.81 \times 10^{-4}$ cm/s$^2$[7] therefore resulting to limit for maximal mass of C/2012 S1 nucleus as $1.24 \times 10^{11}$ kg, almost half as C/2003 T4 but nearly the same as C/1999 S4.

For the period covered by SWAN instrument, the $Q_{H2O}$ of ISON between 24. Oct to 21. Nov fits value 28.302 - 3.84 log(r). Amount of water produced in this interval, calculated from visual magnitudes, is only 7.09 % smaller, so it is possible to correct the equation for comet ISON and use visual magnitudes to calculate amount of produced water in larger interval outside SWAN data. In the time, when comet had smaller heliocentric distance then 1,5 to its disintegration before perihelion, result is $\sim 2.64 \times 10^{10}$ kg water loss, or >21 % fraction of total nucleus mass.

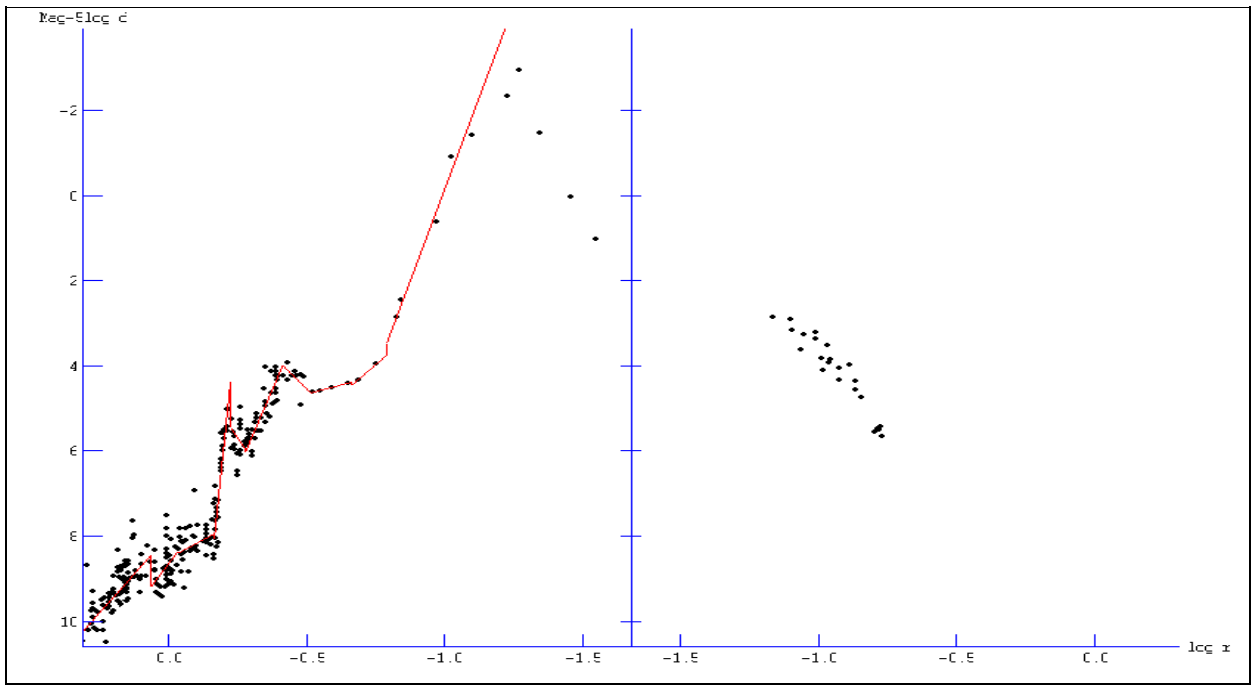

**Figure 3.** Light curve of comet C/2012 S1 (ISON) between 1.5 AU distance from Sun to perihelion. Magnitude evolution is very chaotic and intervals need to be covered by many various photometric parameters.

### 3. Conclusions

Water production rates calculated from visual magnitudes are proved to give us basic view on cometary physical properties.

As both comets C/1999 S4 (LINEAR) and C/2012 S1 (ISON) failed to survive their perihelion passage and the water fraction of their nucleus masses are finite, we can state that comet ISON had nearly three times more water fraction of total nucleus mass. Comet C/2003 T4 (LINEAR) survived its passage around Sun, but loss more than 13 % of its nucleus mass only on water. Its nucleus remains intact however after extensive mass loss, it was much smaller and less active than before perihelion. But as its activity continues far away from Sun, we can conclude that nucleus wasn't exhausted for all water during this revolution around Sun.

## 4. Discussion

Water ice can play major role for cometary nucleus surviving prospect as the sublimating molecules taking away heat from fragile mineral content and therefore works as protection against heat stress. Comet ISON which had 3 times larger water fraction of its nucleus had survived to very small distance 0.024 AU from Sun, while much drier C/1999 S4 undergone final breakup in much larger distance 0.767 AU.

Larger perihelion distance and enough nucleus water content protected C/2003 T4 nucleus from its disintegration, but it sustain heavy mass loss and the next return could be fatal for its further existence, however planetary perturbations will place this comet on hyperbolic orbit, so it should leave our solar system.

Combining expected rate of water production $1\times10^{28}$ molecules $H_2O$ from 1 km$^2$ on nucleus surface per second at 1 AU[9] distance and mean density for calculated masses 0.5 g/cm$^3$ resulting to unrealistic surface active fraction over 100% for spherical nucleus of all 3 comets. Partially this can be caused by more complicated surface topology as nucleus shape is far away from ideal sphere, however other part can be result of water sublimating from icy grains in coma as was previously discovered for several comets[10][11]. Such water amount can't contribute to non-gravitational forces on nucleus, therefore mass of water in equations is smaller., so the nucleus mass must be smaller too then values calculated above and the water fraction of their mass must be larger.